\documentclass{emulateapj}

\newcommand{\nature}{Nature}                    % Nature
\newcommand{\fdis}{Faraday~Discuss.}            % Faraday Discussions
\newcommand{\jpcrd}{J.~Phys.~Chem.~Ref.~Data}   % Journal of Physical and Chemical Reference Data
\shorttitle{Photochemistry in the inner layers of clumpy
circumstellar envelopes} \shortauthors{Ag\'undez et al.}

\begin{document}

\title{Photochemistry in the inner layers of clumpy circumstellar envelopes:\\ formation of water in C-rich objects and of C-bearing molecules in O-rich objects}

\author{Marcelino Ag\'undez\altaffilmark{1}, Jos\'e Cernicharo\altaffilmark{2} and Michel Gu\'elin\altaffilmark{3}}

\altaffiltext{1}{LUTH, Observatoire de Paris-Meudon, 5 Place Jules
Janssen, 92190 Meudon, France; marcelino.agundez@obspm.fr}

\altaffiltext{2}{Departamento de Astrof\'isica, Centro de
Astrobiolog\'ia, CSIC-INTA, Ctra. de Torrej\'on a Ajalvir km 4,
Torrej\'on de Ardoz, 28850 Madrid, Spain}

\altaffiltext{3}{Institut de Radioastronomie Millim\'etrique, 300
rue de la Piscine, 38406 Saint Martin d'H\'eres, France}

\begin{abstract}
A mechanism based on the penetration of interstellar ultraviolet
photons into the inner layers of clumpy circumstellar envelopes
around AGB stars is proposed to explain the non-equilibrium
chemistry observed in such objects. We show through a simple
modelling approach that in circumstellar envelopes with a certain
degree of clumpiness or with moderately low mass loss rates (a few
10$^{-7}$ M$_{\odot}$ yr$^{-1}$) a photochemistry can take place
in the warm and dense inner layers inducing important changes in
the chemical composition. In carbon-rich objects water vapor and
ammonia would be formed with abundances of 10$^{-8}$$-$10$^{-6}$
relative to H$_2$, while in oxygen-rich envelopes ammonia and
carbon-bearing molecules such as HCN and CS would form with
abundances of 10$^{-9}$$-$10$^{-7}$ relative to H$_2$. The
proposed mechanism would explain the recent observation of warm
water vapor in the carbon-rich envelope IRC +10216 with the
Herschel Space Observatory, and predict that H$_2$O should be
detectable in other carbon-rich objects.
\end{abstract}


\section{Introduction}

\begin{figure}[b]
\includegraphics[angle=0,scale=.45]{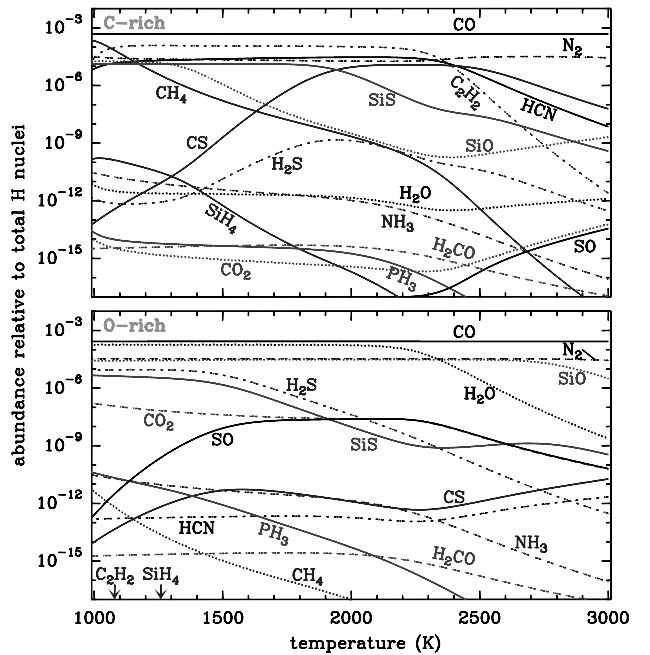}
\caption{TE abundances as a function of temperature for a C- and
O-rich gas (with a C/O abundance ratio of 1.5 and 0.5
respectively) with a constant particle density of 10$^{14}$
cm$^{-3}$.} \label{fig-lte-abun}
\end{figure}

Evolved stars in the asymptotic giant branch (AGB) phase undergo
important mass loss processes that produce extended circumstellar
envelopes (CSEs) composed of dust and molecules. The molecular
composition of CSEs is affected by several processes during the
travel towards the interstellar medium (ISM) but it is originally
established in the stellar atmosphere under thermochemical
equilibrium (TE) conditions and is to a large extent governed by
the C/O abundance ratio. In C-rich stars (C/O $>$ 1) most of the
oxygen is in the form of CO resulting in a carbon-based chemistry
while in O-rich stars (C/O $<$ 1) CO locks most of the carbon
leaving little for other molecules (\citealt{tsu73}; see also
Fig.~\ref{fig-lte-abun}).

Astronomical observations have in the main confirmed this picture,
although with a significant number of discrepancies. HCN emission
is widely observed in O-rich objects \citep{buj94,bie00,dec08}, in
some of them coming from the inner regions of the CSE
\citep{dua00}. Water vapor in the C-rich envelope IRC +10216 is
confined to the warm near surroundings of the star, according to
recent Herschel observations \citep{dec10}. Moreover, ammonia is
observed in the inner regions of C- and O-rich CSEs
\citep{kea93,men95} with abundances much larger than predicted by
TE.

On the theoretical side, shocks induced by the stellar pulsation
have been proposed to explain the non-equilibrium chemistry
observed in the inner regions of CSEs \citep{che06}. These models
explain the formation of HCN and CS in the inner wind of O-rich
CSEs, although fail to explain the presence of H$_2$O in IRC
+10216 and of NH$_3$ in both C- and O-rich CSEs.

In this Letter we investigate an alternative mechanism of
non-equilibrium chemistry, based on the penetration of
interstellar ultraviolet (UV) photons into the inner regions of
CSEs with a certain degree of clumpiness.

\section{Model}

The model is based on a central AGB star surrounded by a spherical
circumstellar envelope, whose physical parameters are given in
Table~\ref{table-phys-model}, and is adopted to investigate the
chemistry in both C- and O-rich CSEs with mass loss rates between
10$^{-7}$ and 10$^{-5}$ M$_{\odot}$ yr$^{-1}$.

In CSEs with an intense mass loss process strictly isotropic and
homogeneous, the inner regions are well shielded from interstellar
UV light, so that only the outer layers are affected by
photochemistry (e.g. \citealt{cha95,wil97}). Observations,
however, have shown that CSEs have usually clumpy structures both
at small and large scales \citep{cha94,gue97,wei98,fon03,lea06},
which allow for a deeper penetration of interstellar UV photons
into the inner regions. To model the effects of clumpiness on the
circumstellar chemistry we adopt a simple approach in which the
CSE consists of two components: a major one whose inner regions
are well shielded against interstellar UV light, and a minor one
(which accounts for a fraction $f_m$ of the total circumstellar
mass) for which the shielding matter located in the radial outward
direction is grouped into clumpy structures leaving a fraction
$f_{\Omega}$ of the solid angle of arrival of interstellar photons
free of matter.

\begin{deluxetable}{lr}
\tabletypesize{\scriptsize} \tablecaption{Model physical
parameters\label{table-phys-model}} \tablewidth{0pc} \startdata
\hline \hline
%Parameter & Value  \\
%\hline
Star effective temperature ($T_*$)        & 2000 K \\
Stellar radius ($R_*$)                    & 5 $\times$ 10$^{13}$ cm \\
Expansion velocity ($v$)$^a$              & 5 km s$^{-1}$ for $r$ $<$ 5 $R_*$ \\
                                          & 15 km s$^{-1}$ for $r$ $\geq$ 5 $R_*$ \\
Gas kinetic temperature ($T$)$^b$         & $T_*$($r$/$R_*$)$^{-0.7}$ \\
Volume density of gas particles ($n$)$^c$ &
$\dot{M}$/(4$\pi$$r^2$$\langle m_{\rm g} \rangle$$v$)
\enddata
\tablecomments{$r$ is the radius measured from the center of the
star. $^a$ Velocity field similar to that adopted in previous
studies of inner CSEs \citep{kea93,fon08}. $^b$ Values of the
exponent are typically between $-0.5$ and $-1$
\citep{jus94,fon08}. $^c$ $\dot{M}$ is the mass loss rate and
$\langle m_{\rm g} \rangle$ is the mean mass of the gas particles
(typically $\sim$ 2.3 times the mass of the hydrogen atom in
CSEs).}
\end{deluxetable}

The gas phase chemical composition of these two components is
computed as they expand from the innermost regions ($r=2 R_*$)
until the end of the CSE. The adopted abundances at the initial
radius are given in Table~\ref{table-initial-abundances}.
$^{13}$CO is also included with an abundance 30 times lower than
$^{12}$CO \citep{mil09}. The chemical network has been used in
previous chemical models of warm and dense UV illuminated regions
\citep{cer04,agu08a}. Photodissociation and photoionization rates
are evaluated as a function of the visual extinction $A_V$
\citep{woo07,van06}, adopting the interstellar UV field of
\citet{dra78}. For $^{12}$CO the photodissociation rate is
evaluated according to \citet{mam88}, who included the effect of
self-shielding, and for $^{13}$CO through the expression 2
$\times$ 10$^{-10}$ $\exp (-2.5 A_V)$ s$^{-1}$ \citep{woo07}.

For the major component, shielded by a smooth envelope, $A_V$
depends on the column density of H nuclei $N_{\rm H}$ in the
radial outward direction as $A_V$ = $N_{\rm H}$(cm$^{-2}$)/1.87
$\times$ 10$^{21}$ \citep{boh78}. The UV field for this component
may be expressed as:
\begin{equation}
4 \pi J_{\lambda} = I_{\lambda} \Omega \exp \Big\{-
\frac{[A_{\lambda}/A_V]}{1.086} A_V \Big\}
\label{eq-uvflux-homogeneous}
\end{equation}
where $I_{\lambda}$ is the unattenuated interstellar UV field at a
wavelength $\lambda$, $\Omega$ the solid angle of arrival of most
of the UV flux (depends strongly on the radius and ranges from a
small fraction of $\pi$ sr in the inner regions up to almost
4$\pi$ sr in the outermost layers), and [$A_{\lambda}$/$A_V$] the
ratio of the dust extinction at $\lambda$ and at visual
wavelengths (3.8 for $\lambda$ = 1250 \AA as found for the ISM by
\citealt{fit90}). For the minor component, shielded by a clumpy
envelope, the UV field may be analogously expressed as:
\begin{equation}
4 \pi J_{\lambda} = I_{\lambda} \Omega \Big[ (1-f_{\Omega}) \exp
\Big\{- \frac{[A_{\lambda}/A_V]}{1.086} A_V \Big\} + f_{\Omega}
\Big] \label{eq-uvflux-clumpy}
\end{equation}
For the minor component we compute an effective visual extinction
$A_V^{\rm ef}$ by substituting $A_V$ by $A_V^{\rm ef}$ into
Eq.~(\ref{eq-uvflux-homogeneous}) and equalling the right parts of
Eqs.~(\ref{eq-uvflux-homogeneous}) and (\ref{eq-uvflux-clumpy}).

Near-infrared interferometry, able to trace the dust emission at
milli-arcsecond scales, has revealed a extremely clumpy structure
in objects such as IRC +10216, with five individual clumps in the
close surroundings of the star, some of them with angle
separations of up to 20$-$30$^\circ$ \citep{wei98}.
Millimeter-wave interferometry of molecular lines can, unlike
infrared observations of dust, provide information on the
projected velocity in the line of sight and allow to build three
dimensional maps. Observations of IRC +10216 in molecular lines of
CN, C$_2$H, C$_4$H, and HC$_5$N \citep{gue93,din08} have shown
that these species are distributed in a circumstellar shell with a
radius of 15-20$''$, and with two conical holes in the NNE and SSW
directions having an aperture angle of about 30$^\circ$. This
corresponds to a solid angle of $\pi$/4 sr, which may be a good
fraction of $\Omega$ in the inner circumstellar regions. Both
$f_m$ and $f_{\Omega}$ are phenomenological parameters, difficult
to quantify in any CSE. Anyway, adopting $f_m$ = 0.1$-$0.2 (the
minor UV illuminated component accounting for just a 10$-$20 \% of
the total circumstellar mass) and $f_{\Omega}$ = 0.2$-$0.5 (in
line with the observational data described above for IRC +10216)
seems reasonable for a CSE with a sufficient degree of clumpiness
and allows for a sufficient penetration of interstellar UV photons
into the inner layers of the CSE. For all the models we have
adopted $f_m$ = 0.1 and $f_{\Omega}$ = 0.25.

\begin{deluxetable}{lrr@{\hspace{1.0cm}}lrr}[hb]
\tabletypesize{\scriptsize} \tablecaption{Initial abundances
relative to H$_2$ in C- and O-rich
CSEs\label{table-initial-abundances}} \tablewidth{0pc} \startdata
\hline \hline
\multicolumn{3}{c}{Carbon-rich} & \multicolumn{3}{c}{Oxygen-rich} \\
\hline
Species    & Abundance              & Ref       & Species    & Abundance              & Ref  \\
\hline
He         & 0.17                   &           & He         & 0.17                   & \\
CO         & 8 $\times$ 10$^{-4}$   & (1)       & CO         & 3 $\times$ 10$^{-4}$   & (1) \\
N$_2$      & 4 $\times$ 10$^{-5}$   & (2)       & N$_2$      & 4 $\times$ 10$^{-5}$   & (2) \\
C$_2$H$_2$ & 8 $\times$ 10$^{-5}$   & (3)       & H$_2$O     & 3 $\times$ 10$^{-4}$   & (4) \\
HCN        & 2 $\times$ 10$^{-5}$   & (3)       & CO$_2$     & 3 $\times$ 10$^{-7}$   & (5) \\
SiO        & 1.2 $\times$ 10$^{-7}$ & (3)       & SiO        & 1.7 $\times$ 10$^{-7}$ & (6) \\
SiS        & 10$^{-6}$              & (3)       & SiS        & 2.7 $\times$ 10$^{-7}$ & (7) \\
CS         & 5 $\times$ 10$^{-7}$   & (3)       & SO         & 10$^{-6}$              & (8) \\
SiC$_2$    & 5 $\times$ 10$^{-8}$   & (3)       & H$_2$S     & 7 $\times$ 10$^{-8}$   & (9) \\
HCP        & 2.5 $\times$ 10$^{-8}$ & (3)       & PO         & 9
$\times$ 10$^{-8}$    & (10)
\enddata
\tablerefs{(1) \citet{tey06}; (2) TE abundance; (3) abundance in
IRC +10216, see \citet{agu09}; (4) \citet{mae08}; (5)
\citet{tsu97}; (6) \citet{sch06}; (7) \citet{sch07}; (8)
\citet{buj94}; (9) \citet{ziu07}; (10) \citet{ten07}.}
\end{deluxetable}

\begin{figure*}
\includegraphics[angle=-90,scale=.69]{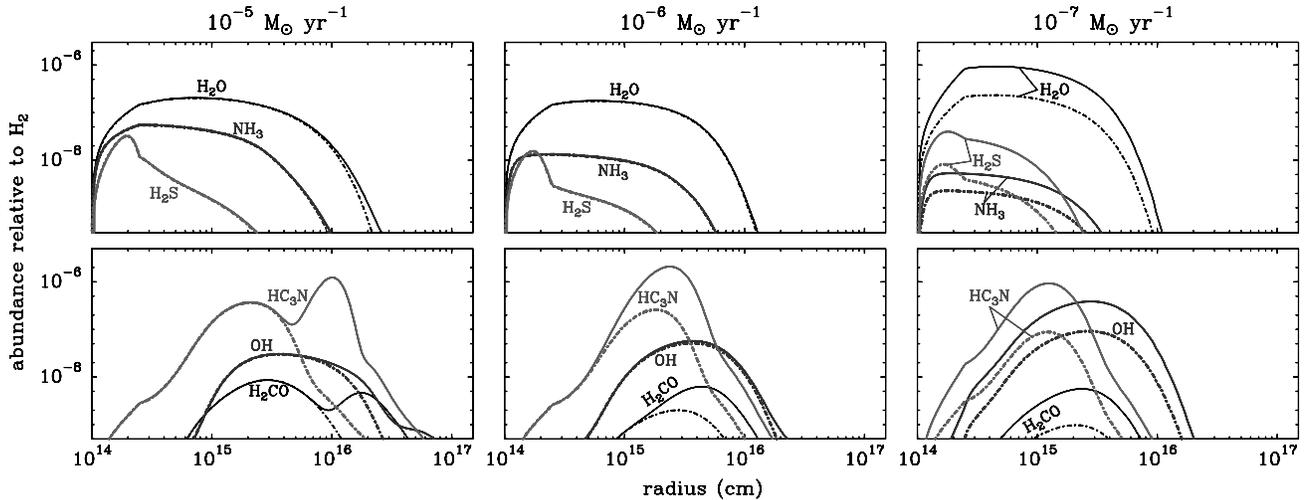}
\caption{Calculated abundance of several molecules as a function
of radius for carbon-rich CSEs with mass loss rates of 10$^{-5}$,
10$^{-6}$, and 10$^{-7}$ M$_{\odot}$ yr$^{-1}$. Dashed-dotted
lines correspond to the abundance of the minor UV exposed
component properly corrected by the factor $f_m$, while continuous
lines correspond to the abundance weighted-averaged over the minor
and major components, which can be expressed as
$\overline{x}_i(r)$ = (1 - $f_m$)$x^M_i(r)$ + $f_m$ $x^m_i(r)$,
where $x^M_i(r)$ and $x^m_i(r)$ are the abundances of the species
$i$ at a radius $r$ in the major and minor components,
respectively. Although in the reality the minor UV exposed and the
major UV shielded components would not coexist in space, this
treatment allows to simplify taking the abundance averaged over
all radial directions. Note that in those regions where the
dashed-dotted and continuous lines of a given species coincide,
only the minor UV exposed component contributes to the abundance.}
\label{fig-crich-chem}
\end{figure*}

\section{Results}

Fig.~\ref{fig-lte-abun} shows the TE abundance of various
molecules calculated under conditions typical of the stellar
atmosphere of AGB stars. The calculations have been done with the
code described in \citet{tej91} and most of the thermochemical
data have been taken from \citet{cha98}. The main purpose of these
calculations is to show that several molecules which are observed
in the inner regions of CSEs have very low TE abundances (e.g.
H$_2$O, NH$_3$, and SiH$_4$ in C-rich objects, and HCN, CS, and
NH$_3$ in O-rich objects).\\

\begin{figure*}
\includegraphics[angle=-90,scale=.69]{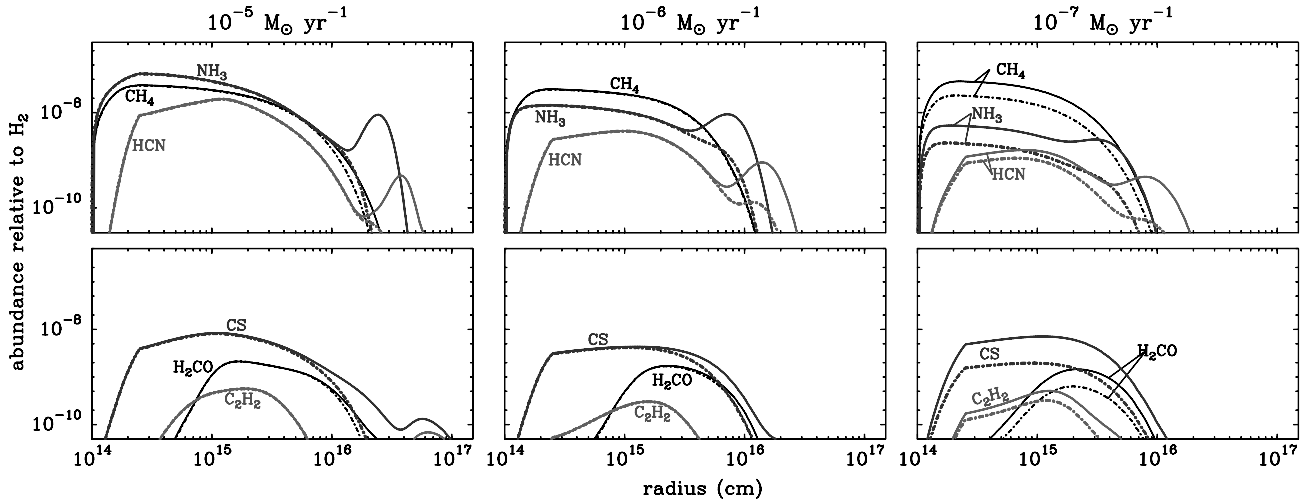}
\caption{Same as in Fig.~\ref{fig-crich-chem} but for oxygen-rich
CSEs.} \label{fig-orich-chem}
\end{figure*}

Now focusing on the models based on chemical kinetics,
Fig.~\ref{fig-crich-chem} and Fig.~\ref{fig-orich-chem} show the
calculated abundance distribution of some molecules in C- and
O-rich CSEs with mass loss rates of 10$^{-5}$, 10$^{-6}$, and
10$^{-7}$ M$_{\odot}$ yr$^{-1}$.

In C-rich CSEs with mass loss rates as high as 10$^{-5}$
M$_{\odot}$ yr$^{-1}$, typical of an object such as IRC +10216,
water vapor would be effectively formed in the dense and warm
inner regions of the minor UV illuminated component, for which
$A_V^{\rm ef}$ is $<$ 1 mag, with a global abundance relative to
H$_2$ in excess of 10$^{-7}$ (see Fig.~\ref{fig-crich-chem}). In
such regions the photodissociation of $^{13}$CO and SiO, the major
reservoirs of oxygen besides $^{12}$CO (hard to photodissociate
due to self-shielding effects), liberates atomic oxygen which is
effectively converted into water through the chemical reactions
\begin{equation}
\rm O + H_2 \rightarrow OH + H
\end{equation}
\begin{equation}
\rm OH + H_2 \rightarrow H_2O + H
\end{equation}
which despite having activation barriers are rapid enough due to
the high temperatures attained in these inner layers. The
photodestruction of molecules in the minor UV illuminated
component occurs fast, but for some of them (e.g. H$_2$O) the
formation rate is high enough to allow them to extend up to
relatively large radii, $\sim$ 10$^{16}$ cm (see
Fig.\ref{fig-crich-chem}). This mechanism would explain the
observation with Herschel of warm water vapor in the inner
circumstellar regions of the carbon star IRC +10216 \citep{dec10}.
Other mechanisms proposed such as sublimation of cometary ices
\citep{mel01}, Fischer-Tropsch catalysis on the surface on iron
grains \citep{wil04}, or radiative association between O and H$_2$
\citep{agu06}, place water in cool regions located farther than
10$^{15}$ cm.

The proposed mechanism would also have some other interesting
consequences. Hydrides other than H$_2$O, such as NH$_3$, CH$_4$,
H$_2$S, SiH$_4$, and PH$_3$, could also be effectively formed in
the inner CSE by successive hydrogenation reactions of the heavy
atom. All them are observed in IRC +10216
\citep{kea93,cer00,agu08b} with abundances which, except for
CH$_4$, are much larger than predicted by TE (see
Fig.~\ref{fig-lte-abun}). Ammonia, for example, is observed in the
inner CSE of IRC +10216 with an abundance relative to H$_2$ of
10$^{-7}$$-$10$^{-6}$ \citep{kea93,has06}, and yet no efficient
formation mechanism has been proposed, apart from the suggestion
that it could be formed on grain surfaces \citep{kea93}. Our model
predicts an effective formation for NH$_3$ and H$_2$S (see
Fig.~\ref{fig-crich-chem}), but not for SiH$_4$ and PH$_3$ (likely
due to the lack of chemical kinetics data for the relevant
hydrogenation reactions). Other molecules such as HC$_3$N increase
their abundance in the inner envelope due to the penetration of
interstellar UV photons (see Fig.~\ref{fig-crich-chem}), something
that has been recently confirmed through observations of IRC
+10216 at $\lambda$ = 0.9 mm with the IRAM 30-m telescope
(\citealt{dec10}; Kahane et al. in preparation).

Still focusing on C-rich sources (see Fig.~\ref{fig-crich-chem}),
if we move toward lower mass loss rates then the whole CSE starts
to be more transparent to interstellar UV photons. For example,
for a mass loss rate as low as 10$^{-7}$ M$_{\odot}$ yr$^{-1}$
most of H$_2$O is formed in the major UV shielded component, which
is no longer shielded as it has a visual extinction $<$ 1 mag.
Thus, for C-rich CSEs with moderately low mass loss rates (up to a
few 10$^{-7}$ M$_{\odot}$ yr$^{-1}$) we should expect a relatively
large H$_2$O abundance even if the CSE is not particularly clumpy,
prediction that should be easily tested with Herschel.

In the case of O-rich CSEs the penetration of interstellar UV
photons into the inner layers has also interesting chemical
effects (see Fig.~\ref{fig-orich-chem}). Among them it is worth
mentioning the formation of NH$_3$, CH$_4$, HCN, and CS in the
inner envelope with abundances relative to H$_2$ in the range
10$^{-8}$$-$10$^{-7}$, i.e. much larger than predicted by TE
calculations (see Fig.~\ref{fig-lte-abun}). The formation of
C-bearing molecules in a dense and warm UV illuminated O-rich gas
has been discussed by \citet{agu08a} in the context of the
chemistry of protoplanetary disks. HCN, CS, and NH$_3$ are
observed in O-rich CSEs with abundances of 10$^{-7}$$-$10$^{-6}$
relative to H$_2$ \citep{buj94,men95,bie00}, which are somewhat
higher than predicted by us. Other mechanisms based on shocks
induced by the stellar pulsation \citep{dua99,che06} predict
fractional abundances for HCN and CS of 10$^{-6}$$-$10$^{-5}$,
which are somewhat higher than observed, but a negligible
abundance for NH$_3$.

% ref to Talbi study on H2 + O, radiative association.
%
%CO and 13CO photodissociation in detached and clumpy CSEs has been
%investigated by \citet{liu97}

\section{Conclusions}

We have shown through a simple modelling approach that in CSEs
envelopes with a certain degree of clumpiness or with moderately
low mass loss rates (a few 10$^{-7}$ M$_{\odot}$ yr$^{-1}$) a
photochemistry can take place in the warm and dense inner layers
inducing important changes in the chemical composition. This
mechanism allows for the formation of H$_2$O and NH$_3$ in C-rich
objects and HCN, CS, and NH$_3$ in O-rich objects, with abundances
much higher than predicted by thermochemical equilibrium but close
to the values typically derived from astronomical observations.
This mechanism explains the recent observation of warm water vapor
in the carbon-rich envelope IRC +10216 with the Herschel Space
Observatory, and predict that H$_2$O should be detectable in other
carbon-rich objects.

\acknowledgments

%The research leading to these results has received funding from
%the [European Community's] Seventh Framework Programme
%([FP7/2007-2013] under grant agreement n° [xxxxxx].
M.A. is supported by a \textit{Marie Curie Intra-European
Individual Fellowship} within the European Community 7th Framework
Programme under grant agreement n$^{\circ}$ 235753.

\end{document}